\documentclass[twocolumn,showpacs,preprintnumbers,amsmath,amssymb,floatfix]{revtex4}
\usepackage{graphicx}% Include figure files
\usepackage{dcolumn}% Align table columns on decimal point
\usepackage{bm}% bold math

\usepackage[normalem]{ulem}  % \sout{old text} for strikeout
\usepackage[dvips]{color}

\begin{document}

%\preprint{APS/123-QED}

\title{Temperature effects in the nuclear isoscaling}

\author{S.R.\ Souza$^{1,2}$}
\author{M.B.\ Tsang$^3$}
\author{B.V.\ Carlson$^{4}$}
\author{R.\ Donangelo$^{1,5}$}
\author{W.G.\ Lynch$^3$}
\author{A.W.\ Steiner$^3$}
\affiliation{$^1$Instituto de F\'\i sica, Universidade Federal do Rio de Janeiro
Cidade Universit\'aria, \\CP 68528, 21941-972, Rio de Janeiro, Brazil}
\affiliation{$^2$Instituto de F\'\i sica, Universidade Federal do Rio Grande do Sul\\
Av. Bento Gon\c calves 9500, CP 15051, 91501-970, Porto Alegre, Brazil}
\affiliation{$^3$ Joint Institute for Nuclear Astrophysics, National
Superconducting Cyclotron Laboratory, and the Department of Physics
and Astronomy, Michigan State University,
East Lansing, MI 48824, USA}
\affiliation{$^4$Departamento de F\'\i sica, 
Instituto Tecnol\'ogico de Aeron\'autica - CTA, 12228-900\\
S\~ao Jos\'e dos Campos, Brazil}
\affiliation{$^5$Instituto de F\'\i sica, 
Facultad de Ingenier\'\i a, Universidad de la Rep\'ublica,\\
Julio Herrera y Reissig 565, 11.300 Montevideo, Uruguay}

\date{\today}% It is always \today, today,

\begin{abstract}
The properties of the nuclear isoscaling at finite temperature are investigated and the extent to which its parameter $\alpha$ holds
information on the symmetry energy is examined.
We show that, although finite temperature effects invalidate the analytical formulae that relate the isoscaling parameter $\alpha$ to
those of the mass formula, the symmetry energy remains the main ingredient that dictates the behavior of $\alpha$ at
finite temperatures, even for very different sources.
This conclusion is not obvious as it is not true in the vanishing temperature limit, where analytical formulae are available.
Our results also reveal that different statistical ensembles lead to essentially the same conclusions based on the isoscaling analysis,
for the temperatures usually assumed in theoretical calculations in the nuclear multifragmentation process.
\end{abstract}

\pacs{25.70.Pq, 24.60.-k}% PACS, the Physics and Astronomy
\maketitle

\section{\label{sec:introuduction}Introduction\protect}
The scaling law obeyed by the ratio of the yields $Y_{A,Z}(i)$ of fragments of mass and atomic numbers
$A$ and $Z$ observed in different reactions, labeled `1' and `2', {\it i.e.} the nuclear isoscaling
\cite{isoscaling1,isoscaling2}

\begin{equation}
R_{21}=\frac{Y_{A,Z}(2)}{Y_{A,Z}(1)}=C\exp(\alpha N+\beta Z)\;,
\label{eq:iso}
\end{equation}

\noindent
opens the possibility of accessing the properties of nuclear matter far from equilibrium.
The parameter $C$ in the above Eq.\ is just a normalization factor, but $\alpha$ and $\beta$ hold
valuable information on the nuclear interaction \cite{isoscaling3,isoscWolfgangBotvina1}.
The relevance of this property is due to the fact that, if the sources formed in the
intermediate stages of the two reactions have the same temperature $T$, effects associated
with the decay of the primary hot fragments should not distort this scaling law.
Since it is difficult to tune the temperature of the sources experimentally, similar reactions are
usually used so that one expects $T$ to be very close in both sources.
Therefore, the parameters $\alpha$ and $\beta$ preserve information related to the stages
at which the fragments are created.
This assumption is supported by theoretical calculations \cite{isoscaling3,isocc,isoSymmetryBotvina2006}.

In the vanishing temperature limit and for similar sources, it has been shown that the symmetry energy coefficient of the nuclear
mass formula, $C_{\rm sym}$, gives the main contribution to the parameter $\alpha$ \cite{isoscaling3,isoscWolfgangBotvina1}

\begin{equation}
\alpha=4\frac{C_{\rm sym}}{T}\left[\left(\frac{Z_1}{A_1}\right)^2-\left(\frac{Z_2}{A_2}\right)^2\right]\;,
\label{eq:isoAlpha}
\end{equation}

\noindent
where $A_i$ and $Z_i$ stand for the mass and atomic numbers of the sources formed in the two reactions.
This close connection between $\alpha$ and $C_{\rm sym}$ has extensively been exploited in
many studies \cite{isoscalingIndraGSI2005,isoSymmetryBotvina2006,SouliotisCsym2009,symEnergySamaddar2008,symEnergyShatty2007,
BotvinaSurf2006,isoNatowitz2007_2,isoSouliotis2007,IsospinSymmetry,RadutaIsoSym1,RadutaIsoSym2,isoMassFormula2008,
isoscalingAMD2003,symEnergyBaoAnLi2006,symEnergyBaoAnLi2007}.

The density dependence of $C_{\rm sym}$ is of particular interest \cite{isoscalingIndraGSI2005,isoSymmetryBotvina2006,symEnergyShatty2007,SouliotisCsym2009,symEnergySamaddar2008,symEnergyBaoAnLi2006,symEnergyBaoAnLi2007}.
Since $\alpha$ is expected to retain the memory from the breakup configuration, it could provide relevant information
on the freeze-out stage.
However, the formal derivation of the relationship between $\alpha$ and $C_{\rm sym}$, Eq.\ (\ref{eq:isoAlpha}), has been carried out
in the framework of statistical models, in which the created fragments are at the saturation density
\cite{isoscaling3,isoscWolfgangBotvina1}.
As a matter of fact, it has been only very recently that the thermal expansion of the fragments at the freeze-out stage has been
consistently incorporated in the statistical models \cite{smmtf1}.
Furthermore, although the derivation presented in Ref.\ \cite{isoscWolfgangBotvina1} clearly shows that the leading term relating
$\alpha$ to $C_{\rm sym}$ is indeed given by Eq.\ (\ref{eq:isoAlpha}), it assumed that the temperature of the
decaying source tends to zero.
Since in this temperature range the system's density must be close to the saturation value, it is not clear whether the
conclusions implied by this Eq.\ should still hold at higher temperatures, where effects associated with the multifragment
emission are non-negligible.
Nevertheless, a strict interpretation of this expression has been adopted in many works
\cite{isoscalingIndraGSI2005,isoSymmetryBotvina2006,symEnergyShatty2007,SouliotisCsym2009} to explain the behavior of $\alpha$ at
relatively high temperatures, which led to controversial interpretations based on different scenarios
\cite{symEnergySamaddar2008,RadutaIsoSym1,RadutaIsoSym2,isoMassFormula2008}.

In this work we study the consistency of Eq.\ (\ref{eq:isoAlpha}) using the traditional Statistical Multifragmentation Model (SMM),
in which the fragments' densities at the freeze-out configuration correspond to their saturation value \cite{smm1,smm2,smm4}.
In the context of this model, the expanded breakup volume of the system is associated with the separation between the fragments, rather than
to the thermal dilatation of their volumes.
Therefore, in this framework, the symmetry energy is strictly insensitive to the freeze-out volume of the system.

We investigate whether the conclusions based on Eq.\ (\ref{eq:isoAlpha}), formally derived in the vanishing temperature limit
\cite{isoscWolfgangBotvina1}, remain valid in the temperature range usually assumed in the nuclear multifragmentation process.
Although there are no analytic formulae relating $\alpha$ to $C_{\rm sym}$ at non-vanishing temperatures,
we show numerically that the symmetry energy rules the behavior of $\alpha$ at these temperatures.
This conclusion remain valid also in the case where the other terms of the mass formula, disregarded in the derivation of
Eq.\ (\ref{eq:isoAlpha}), dominate its behavior at low temperatures.
We demonstrate that the main conclusions based on this expression still hold although Eq.\ (\ref{eq:isoAlpha}) breaks down in this
temperature range.
More specifically, we show that, contrary to what is expected from Eq.\ (\ref{eq:isoAlpha}),
$C_{\rm sym}\ne T\alpha/4[(Z_1/A_1)^2-(Z_2/A_2)^2]$, except when $T\rightarrow 0$.
It should be noticed that even if former studies 
\cite{isoscaling3,isoscWolfgangBotvina1,isoscalingIndraGSI2005,isoSymmetryBotvina2006,symEnergySamaddar2008,symEnergyShatty2007,
BotvinaSurf2006,isoNatowitz2007_2,isoSouliotis2007,IsospinSymmetry,RadutaIsoSym1,RadutaIsoSym2,isoMassFormula2008,
isoscalingAMD2003,symEnergyBaoAnLi2006,symEnergyBaoAnLi2007}
clearly show that the isoscaling parameter $\alpha$ and $C_{\rm  sym}$
are strongly correlated, the extent to which the functional dependence given by Eq.\ (\ref{eq:isoAlpha}) is a good approximation
to the actual relationship between $\alpha$ and $C_{\rm sym}$ is not clear.

The sensitivity of the isoscaling analysis to the statistical ensemble employed in the calculation
is also investigated and we demonstrate that,
in the temperature range relevant to nuclear multifragmentation, the micro-canonical (M.C.), canonical and the grand-canonical (G.C.)
ensembles lead essentially to the same conclusions.

The remainder of the paper is organized as follows.
We briefly recall the main features of the SMM model in sect.\ \ref{sec:model}. 
The results are presented in sect.\ \ref{sec:results}.
We summarize in sect.\ \ref{sec:conclusions} what we have learned.

\section{\label{sec:model}Theoretical framework\protect}
The Helmholtz free energy $F$ is the main ingredient of statistical models since, besides allowing for the calculation of the
thermodynamical properties of the system, it holds the physical ingredients used in the model.
In the SMM, $F$ is written as:

\begin{eqnarray}
\label{eq:fe}
&& F(T)=\frac{C_{\rm c}}{(1+\chi)^{1/3}}\frac{Z_0^2}{A_0^{1/3}}+F_{\rm trans}(T)\\
&&+\sum_{A,Z}N_{A,Z}\left[-B_{A,Z}+f^*_{A,Z}(T)-\frac{C_{\rm c}}{(1+\chi)^{1/3}}\frac{Z^2}{A^{1/3}}\right]\nonumber
\end{eqnarray}

\noindent
where

\begin{eqnarray}
 F_{\rm trans}=-T(M-1)\log\left(V_f/\lambda_T^3\right)+T\log\left(g_0A_0^{3/2}\right)&&\\
 -T\sum_{A,Z}N_{A,Z}\left[\log\left(g_{A,Z}A^{3/2}\right)-\frac{1}{N_{A,Z}}\log(N_{A,Z}!)\right]\;.&&\nonumber
\label{eq:fekin}
\end{eqnarray}

\noindent
In the above expressions, $V_\chi/V_0=1+\chi$ denotes the ratio of the freeze-out volume $V_\chi$ to the ground state volume
$V_0$ of the source, $A_0$ and $Z_0$ are its mass and atomic numbers, respectively.
In this work, we use $\chi=2$ in all the calculations.
The fragment multiplicity in each fragmentation mode is represented by $N_{A,Z}$, $V_f=V_\chi-V_0$ is the free volume,
$\lambda_T=\sqrt{2\pi\hbar^2/mT}$ corresponds to the thermal wavelength, where $m$ is the nucleon mass.
For $A \le 4$ empirical values for the spin degeneracy factor $g_{A,Z}$ are used as well as for the binding energy $B_{A,Z}$.
These light fragments, except for the alpha particle, are assumed to behave like point particles with no internal degrees of
freedom.
In this work, we adopt the simple Liquid Drop Mass (LDM) formula used in Ref.\ \cite{isoMassFormula2008} to calculate the binding
energies of heavier nuclei:

\begin{equation}
B_{A,Z}=C_v A - C_s A^{2/3} -C_c\frac{Z^2}{A^{1/3}}+C_d\frac{Z^2}{A}\;,
\label{eq:be}
\end{equation}

\noindent
where

\begin{equation}
C_i=a_i\left[1-k_i\left(\frac{A-2Z}{A}\right)^2\right]
\label{eq:ci}
\end{equation}

\noindent
and $i=v,s$ corresponds to the volume and surface terms, respectively.
For the sake of clarity in the subsequent calculations discussed in sect.\ \ref{sec:results}, we have suppressed the pairing term.
The coefficient $C_c$ corresponds to the usual Coulomb term, whereas corrections associated with the surface diffuseness are taken
into account by the factor proportional to $C_d$.
Two simple versions of this LDM formula are used below, labeled LDM1 and LDM2 as in Ref.\ \cite{isoMassFormula2008}.
In the first one, $k_s=0$ so that surface corrections to the symmetry energy are neglected and one has
$E^{(1)}_{\rm sym}=C_{\rm sym}(A-2Z)^2/A$, $C_{\rm sym}=a_v k_v$.
The Coulomb correction proportional to $C_d$ is also suppressed in the LDM1 formula, {\it i.e.} $C_d=0$.
In the case of the LDM2 these terms are preserved and one would have 
$E^{(2)}_{\rm sym}=\left[C_{\rm sym}-a_sk_s/A^{1/3}\right](A-2Z)^2/A\;.$
For the sake of the comparisons below, we keep the definition of $C_{\rm sym}$ used in the LDM1 and do not include the
term $-a_sk_s/A^{1/3}$.

The internal free energy of the fragment $f^*_{A,Z}(T)$ is given by the standard SMM expression \cite{smm1}:

\begin{equation}
f^*_{A,Z}(T)=-\frac{T^2}{\epsilon_0}A+\beta_0A^{3/2}\left[\left(\frac{T_c^2-T^2}{T_c^2+T^2}\right)^{5/4}-1\right]\;,
\label{eq:fein}
\end{equation}

\noindent
where $\beta_0=18.0$~MeV, $T_c=18.0$~MeV, and $\epsilon_0=16.0$~MeV.
This formula is used for all nuclei with $A>4$.
In the case of the alpha particles, we set $\beta_0=0$ in order to take into account, to some extent, the large gap between its
ground state and the first excited state.
The spin degeneracy factor for these nuclei is set to unity, due to the schematic treatment of their excited states.

We stress that, since the parameters of the mass formula are temperature and density independent, the symmetry energy
of the fragments are not sensitive to the freeze-out density of the source.
Furthermore, since in the version of SMM employed in this work the internal free energy of the fragments, Eq.\ (\ref{eq:fein}), is
isospin independent, the entropic terms of the Helmholtz free energy do not lead to any contribution to the symmetry energy.

In the translational contribution to the free energy $F_{\rm trans}$, the factor $M-1$ (rather than $M=\sum_{A,Z}N_{A,Z}$) and
$T\log\left(g_0A_0^{3/2}\right)$ arise due to the subtraction of the center of mass motion.

These ingredients are used in all the different ensembles employed in this work, {\it i.e.}, micro-canonical, canonical,
and grand-canonical.
For practical reasons, as it allows for extremely fast calculations,
we used the McGill version of the canonical ensemble of SMM \cite{BettyPhysRep2005}.
Since we adopt the same ingredients, it is equivalent to the traditional canonical Monte Carlo version of SMM \cite{smmIsobaric}.
We refer the reader to Ref. \cite{isoMassFormula2008} for details on the grand-canonical ensemble calculations.

\section{\label{sec:results}Results and discussion\protect}

As has been demonstrated several times \cite{isoscaling3,isoscWolfgangBotvina1,isocc},
the isoscaling parameter $\alpha$ is associated with the baryon chemical potentials $\mu^{(i)}_B$, $i=1,2$, of the sources $S_1$ and 
$S_2$ through

\begin{equation}
\alpha=\frac{\mu_B^{(2)}-\mu_B^{(1)}}{T}\;.
\label{eq:alphaMu}
\end{equation}

\noindent
This relationship is obtained in a scenario in which the fragments' densities correspond
to the saturation value and, therefore, simple analytical expressions for the average yields can be
obtained \cite{isoMassFormula2008}.
Otherwise, the free volume dependence on the fragments' multiplicities \cite{smmtf1} lead to highly non-linear terms in the Helmholtz
free-energy.
As a consequence, the average yields are no longer given by the simple traditional grand-canonical formulae.
Since in this work the fragments' densities are fixed and correspond to their ground state values, these difficulties can be
disregarded.
One of the underlying assumptions made in the derivation of Eq.\ (\ref{eq:alphaMu}) is that the two sources have the same temperature.

The chemical potentials are strongly influenced by the LDM employed in the calculation as well as by $T$ and the composition
of the source.
In the vanishing temperature limit, it has been demonstrated \cite{isoscWolfgangBotvina1}
for the LDM1 that

\begin{equation}
\mu_B^{(i)}(T\rightarrow 0)=-a_v+\frac{a_s}{A_i^{1/3}}-C'_c\frac{Z_i^2}{A_i^{4/3}}
+C_{\rm sym}\left[1-\frac{4Z_i^2}{A_i^2}\right]\;,
\label{eq:muLDM}
\end{equation}

\noindent
where $C'_c=C_c[1-1/(1+\chi)^{1/3}]$.
It is easy to extend this formula to the LDM2, also in the $T\rightarrow 0$ limit, and from Eq.\ (\ref{eq:alphaMu}) one obtains:

\begin{eqnarray}
T\alpha&=&C_{\rm sym}\Delta(Z/A)^2
+a_s\left[\frac{1}{A_2^{1/3}}-\frac{1}{A_1^{1/3}}\right]\nonumber\\
&-&\frac{k_sa_s}{A_2^{1/3}}\left[1-\frac{4Z_2^2}{A_2^2}\right]
+\frac{k_sa_s}{A_1^{1/3}}\left[1-\frac{4Z_1^2}{A_1^2}\right]\nonumber\\
&-& C'_c\left[\frac{Z_2^2}{A_2^{4/3}}-\frac{Z_1^2}{A_1^{4/3}}\right]+C_d\frac{\Delta (A/Z)^2}{4}\;.
\label{eq:alphaT}
\end{eqnarray}

\noindent
In the above expression, we have introduced the shorthand notation $\Delta(Z/A)^2\equiv 4(Z_1^2/A_1^2-Z_2^2/A_2^2)$.

It should be noted that in the statistical models the properties of the sources are taken for granted, as the
dynamics which leads to the configuration at the freeze-out stage is beyond the scope of the approach.
Therefore, we explicitly assume that the sources $S_1$ and $S_2$ have the same temperature $T$.
The validity of this assumption must be investigated on the basis of dynamical models.
For the present purpose, which consists in verifying the consistency of Eq.\ (\ref{eq:isoAlpha}) with the statistical treatment
upon which this expression is based, the pair of sources ($S_1$,$S_2$) are conveniently selected according to their $Z/A$ ratio, as
well as their sizes.

If the mass and atomic numbers of the sources are similar, the symmetry energy dominates the sum and, by neglecting factors involving
$a_s$, $a_sk_s$, as well as the Coulomb terms, it leads to Eq.\ (\ref{eq:isoAlpha}), usually employed in the literature.
From this, one may obtain the apparent symmetry energy coefficient:

\begin{equation}
C_{\rm sym}^{\rm Unc}=T\alpha/\Delta (Z/A)^2\;,
\label{eq:CsymU}
\end{equation}

\noindent
which we label as ``Unc'' to emphasize that the corrections associated with the remaining terms of the mass formula have been discarded.

If the sources $S_1$ and $S_2$ are not similar enough, {\it i.e.} the differences of their mass and atomic numbers are not small,
the corrections neglected in the above expression introduce a fixed shift to $C_{\rm sym}$ for all temperatures, whose value
depends on the sources considered:

\begin{eqnarray}
&& C_{\rm sym}^{\rm Cor}=\Big\{T\alpha
+ C'_c\left[\frac{Z_2^2}{A_2^{4/3}}-\frac{Z_1^2}{A_1^{4/3}}\right]
-C_d\frac{\Delta (Z/A)^2}{4}\nonumber\\
&&+\frac{k_sa_s}{A_2^{1/3}}\left[1-\frac{4Z_2^2}{A_2^2}\right]
-\frac{k_sa_s}{A_1^{1/3}}\left[1-\frac{4Z_1^2}{A_1^2}\right]\nonumber\\
&& -a_s\left[A_2^{-1/3}-A_1^{-1/3}\right]\Big\}/\Delta (Z/A)^2
\label{eq:CsymC}
\end{eqnarray}

\noindent
and the label ``Cor'' denotes this fact.

\begin{figure}[t]
\includegraphics[width=7.8cm]{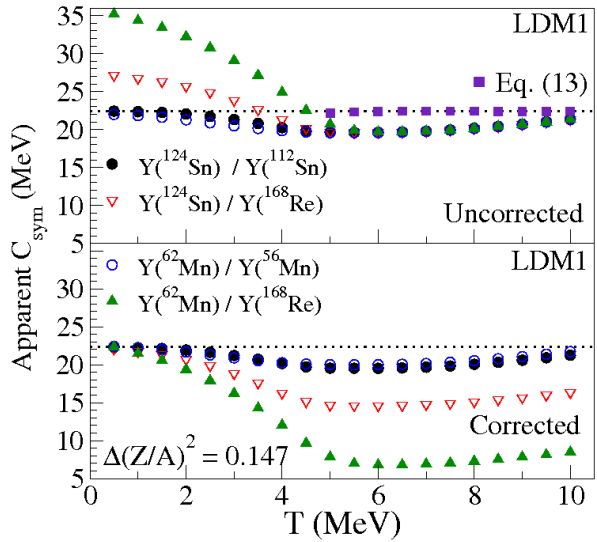}
\caption{\label{fig:isoLDM1} (Color online) Apparent symmetry energy coefficient $C_{\rm sym}^{\rm Unc/Cor}$ for different systems
(with $\Delta (Z/A)^2=0.147$) calculated with the G.C.
ensemble for the LDM1 formula.
The dashed lines correspond  to $a_vk_v=22.39$~MeV.
For details, see the text.
}
\end{figure}

Since the above expressions, Eqs.\ (\ref{eq:CsymU})-(\ref{eq:CsymC}), rely on the baryon chemical potential they must be
accurate only in the $T\rightarrow 0$ limit.
Therefore, it is not clear whether the conclusions implied by any of them remain valid at finite temperatures.
The extent to which $\alpha$ remains related to $C_{\rm sym}$ at these temperatures is not a priori obvious.
It should be observed that studies based on the statistical
\cite{isoscaling3,isoscWolfgangBotvina1,isoscalingIndraGSI2005,isoSymmetryBotvina2006,symEnergySamaddar2008,symEnergyShatty2007,
BotvinaSurf2006,isoNatowitz2007_2,isoSouliotis2007,IsospinSymmetry,RadutaIsoSym1,RadutaIsoSym2,isoMassFormula2008}
and dynamical \cite{isoscalingAMD2003,symEnergyBaoAnLi2006,symEnergyBaoAnLi2007} models have only shown that $\alpha$ is correlated to
$C_{\rm sym}$ but the accuracy of Eq.\ (\ref{eq:isoAlpha}) remains unknown.
Since there are no analytical expressions for $\mu_B$ at finite temperatures, for configurations in which many fragments are present,
this issue is numerically investigated below in the framework of the SMM.

In this way, we reconstruct $C_{\rm sym}^{\rm Unc/Cor}$ through Eqs.\ (\ref{eq:CsymU})-(\ref{eq:CsymC}) and check whether it is
consistent with the fixed parameters used in the LDM adopted in the model.
The parameter $\alpha$ is calculated from fits based on Eq.\ (\ref{eq:iso}) by the yields provided by
the G.C. ensemble (except where stated otherwise) in all the calculations presented below.
If either of the above Eqs.\ is valid at finite temperatures, by construction, one should obtain a constant value
$C_{\rm sym}=a_vk_v$ for all $T$.
Indeed, since the low density values used in the present SMM models is associated with the spacing among the fragments, rather than with
their thermal volume expansion, the reconstructed $C_{\rm sym}^{\rm Unc/Cor}$
should be constant and it should be given by $a_v k_v$ for any temperature and selected sources,
as these parameters are fixed at the saturation density value for all nuclei \cite{ISMMmass,isoMassFormula2008}.
Therefore, deviations from the constant value $a_v k_v$ signal the breakdown of the exact relationship between $\alpha$ and
$C_{\rm sym}$ implied by Eqs.\ (\ref{eq:CsymU})-(\ref{eq:CsymC}) and usually assumed in isoscaling analyses.

Figure \ref{fig:isoLDM1} displays the apparent $C_{\rm sym}^{\rm Unc/Cor}$ as a function of the temperature, obtained with the expressions
above, for the LDM1 ($k_s=0$) and the ($^{112}$Sn,$^{124}$Sn), ($^{168}$Re,$^{124}$Sn), ($^{56}$Mn,$^{62}$Mn), and ($^{168}$Re,$^{62}$Mn)
pairs of sources ($S_1$,$S_2$).
One should notice that $\Delta(Z/A)^2$ is fixed for these sources, so
that the main contribution associated with the symmetry energy is the same in all the cases.
The results corresponding to Eq.\ (\ref{eq:CsymU}) are displayed in the upper panel of this picture and
are very similar for the ($^{112}$Sn,$^{124}$Sn) and ($^{56}$Mn,$^{62}$Mn) sources.
As expected, $C_{\rm sym}^{\rm Unc}$ approaches $a_vk_v=22.39$~MeV at low temperatures, but
important deviations are observed in this limit for the ($^{168}$Re,$^{124}$Sn) and ($^{168}$Re,$^{62}$Mn) sources.
This is because the contributions associated with $a_s$ and the Coulomb terms in Eq.\ (\ref{eq:alphaT}) 
cannot be disregarded in these cases.
The results displayed at the bottom panel of Fig.\ \ref{fig:isoLDM1}, $C_{\rm sym}^{\rm Cor}$, show that the expected value of
$C_{\rm sym}$ is obtained in the low temperature limit when the corresponding shifts, Eq.\ (\ref{eq:CsymC}), are taken into account.

\begin{figure}[t]
\includegraphics[width=7.9cm]{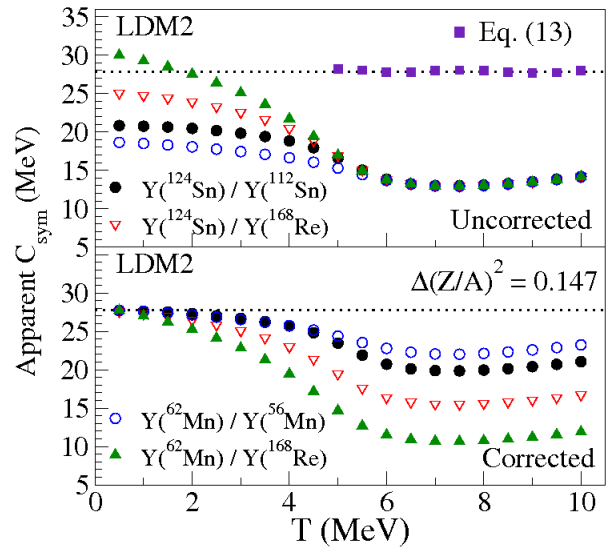}
\caption{\label{fig:isoLDM2} (Color online) Same as Fig.\ \ref{fig:isoLDM1} for the LDM2 formula.
The dashed lines correspond  to $a_vk_v=27.80$~MeV.
For details, see the text.
}
\end{figure}

One striking feature of the upper panel of Fig.\ \ref{fig:isoLDM1} is that, in spite of the large differences observed at low
temperatures between the ($^{168}$Re,$^{124}$Sn) or ($^{168}$Re,$^{62}$Mn) sources and the others, the four curves merge for $T>5.0$~MeV.
This means that Eqs.\ (\ref{eq:CsymU}) and (\ref{eq:CsymC}) break down at large temperatures and $\alpha$
can no longer be approximated by Eq.\ (\ref{eq:isoAlpha}) or (\ref{eq:alphaT}).
However, the fact that the curves given by Eq.\ (\ref{eq:CsymU}) agree for $T>5.0$~MeV implies that, although this Eq.\ is not valid in this
regime, $\alpha$ is indeed strongly correlated with
$C_{\rm sym}$, despite the fact that the analytical temperature dependence of $\alpha$ is unknown.
Furthermore, it also shows that the Coulomb and surface terms play a negligible role for $T>5.0$~MeV,
although they cannot be neglected at smaller temperatures.
Since the nuclear multifragmentation process is expected to take place from this temperature value on, these results suggest that
$\alpha$ remains a valuable observable in these studies.

\begin{figure}[t]
\includegraphics[width=7.9cm]{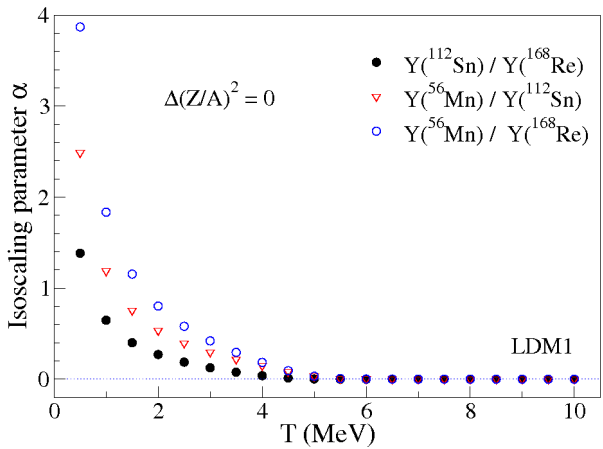}
\caption{\label{fig:alphaZero} (Color online) Isoscaling parameter $\alpha$ calculated with the LDM1 formula.
For details, see the text.
}
\end{figure}

In order to check whether these conclusions are not biased by the simple form of the LDM1 formula, we also performed
calculations using the LDM2 model.
The results are shown in Fig.\ \ref{fig:isoLDM2}.
One sees that the differences are much more pronounced in this case at low temperatures but the qualitative conclusions
above remain true.
For $T>5.0$~MeV, $C_{\rm sym}^{\rm Unc}$ is the same for all the systems, in spite of the very large
differences in the vanishing temperature limit.
This suggests that our findings will not be affected by the use of more involved mass formulae.

The above results show that the relationship between $\alpha$ and $C_{\rm sym}$ is much more complex than 
is suggested by Eqs. ({\ref{eq:isoAlpha}) or ({\ref{eq:alphaT}).
The constraints to which the chemical potentials are submitted at large temperatures, where many fragments are present,
lead to important changes to Eq.\ (\ref{eq:muLDM}), on which these Eqs.\ rely.
In the particular case of the SMM, these results suggest that

\begin{figure}[t]
\includegraphics[width=7.9cm]{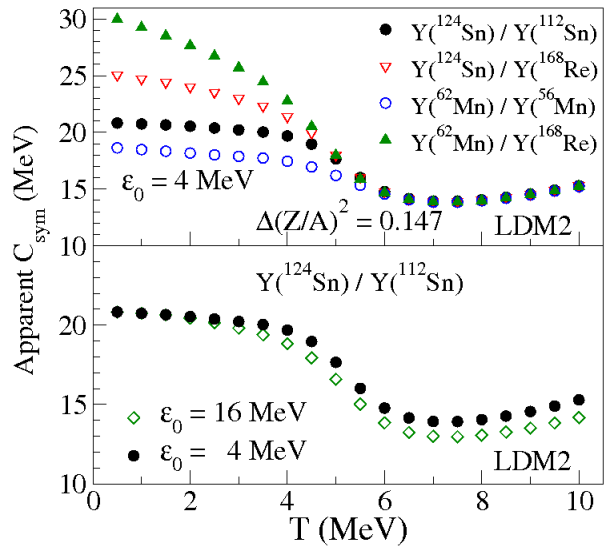}
\caption{\label{fig:CsymLD} (Color online) Uncorrected apparent symmetry energy coefficient $C_{\rm sym}^{\rm Unc}$
calculated from the LDM2 formula for $\epsilon_0=4$~MeV (upper panel).
The bottom panel shows a comparison between the results obtained with $\epsilon_0=4$~MeV and $\epsilon_0=16$~MeV.
For details, see the text.
}
\end{figure}

\begin{eqnarray}
\label{eq:alphaCsymT}
\alpha&=&4\frac{C_{\rm sym}}{T}\left[(Z_1/A_1)^2-(Z_2/A_2)^2\right]\\
&\times & a_0\left(1+a_1T+a_2T^2+a_3T^3\right)\;.\;\;\;\;(T > 5.0\;{\rm MeV})\nonumber
\end{eqnarray}

\noindent
For the LDM1, one has $a_0=1.3338$, $a_1=-0.131842$ MeV$^{-1}$, $a_2=0.0156409$ MeV$^{-2}$, and $a_3=-0.000531862$ MeV$^{-3}$,
whereas $a_0=2.6431$, $a_1=-0.292421$ MeV$^{-1}$, $a_2=0.0338321$ MeV$^{-2}$, and $a_3=-0.00126698$ MeV$^{-3}$ in the case
of the LDM2.
The values of $C_{\rm sym}$ reconstructed from this expression are depicted by the squares in the upper panel of Figs.\
\ref{fig:isoLDM1} and \ref{fig:isoLDM2}.
The results show that $C_{\rm sym}$ is fairly constant and that its value is consistent with the parameters $k_va_v$ employed
in the LDM models.
However, the form of the above expression has been chosen for illustration purposes and the actual relationship between
$\alpha$ and $C_{\rm sym}$ remains unknown.
To obtain a closed expression, one must be able to derive analytical expressions for the chemical potentials associated with
a many fragment system at finite temperature, which has not been achieved yet.

To see that the symmetry energy is, in fact, the essential ingredient that dictates the behavior of the isoscaling parameter $\alpha$
at high temperatures, we consider the ($^{168}$Re,$^{112}$Sn), ($^{112}$Sn,$^{56}$Mn), and ($^{168}$Re,$^{56}$Mn) sources,
for which $\Delta (Z/A)^2=0$.
Equation\ (\ref{eq:alphaT}) predicts that $\alpha\ne 0$ due to contributions from the Coulomb and surface terms, whose
relevance increases as the differences between the mass and/or the atomic numbers of the sources become larger.
The results displayed in Fig. \ref{fig:alphaZero} show that this is true only for $T\rightarrow 0$.
The effect of these terms vanishes at high temperatures and $\alpha\rightarrow 0$ for $T\gtrsim 5.0$~MeV, {\it i.e.}
one sees once more that the Coulomb and surface terms have very little influence on the behavior of $\alpha$ at high temperatures.
The symmetry energy is the key ingredient in this temperature domain, despite the formulae derived at $T\rightarrow 0$.

Since there are no analytical expressions for $\alpha$ at finite temperatures, we investigate the influence of
the internal excitation terms of the free energy by increasing the level density of the bulk term of Eq.\ (\ref{eq:fein}) and
we set $\epsilon_0=4$~MeV.
One should notice that this leads to an important increase in the magnitude of $f^*(T)$.
For instance, for $A=100$ and $T=6.0$~MeV, the standard value is $f^*(T)=-319.4$~MeV whereas for $\epsilon_0=4$~MeV one
has $f^*(T)=-994.4$~MeV.
The apparent $C_{\rm sym}^{\rm Unc}$ is shown in the upper panel in Fig.\ \ref{fig:CsymLD} for $\epsilon_0=4$~MeV and the
LDM2 formula.
One sees that the qualitative features observed previously remain the same and that all the curves merge for $T\gtrsim 5.0$~MeV.
Furthermore, the results displayed in the bottom panel of this picture show that, in spite of the large differences in
$f^*(T)$, the effect on the apparent $C_{\rm sym}^{\rm Unc}$ is very small.
This reveals a fair insensitivity of $\alpha$ to the entropic terms of the internal free energy of the nuclei.
We have checked that similar results are obtained if one reduces $\beta_0$ in Eq.\ (\ref{eq:fein}), which gives the
surface contribution to $f^*(T)$.

\begin{figure}[t]
\includegraphics[width=7.9cm]{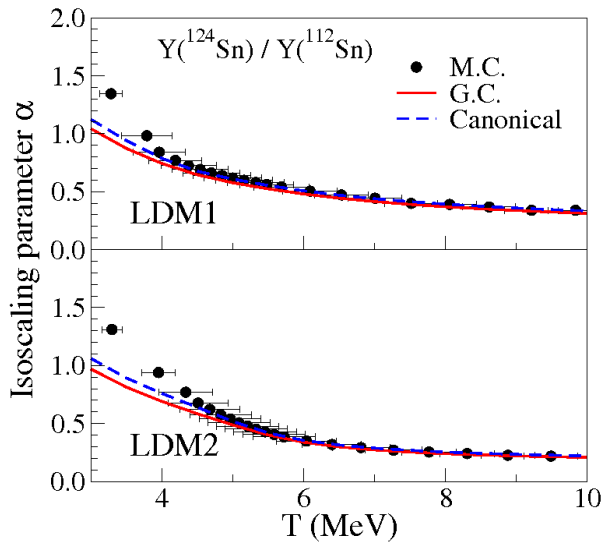}
\caption{\label{fig:alphaEnsembs} (Color online) Isoscaling parameter $\alpha$ calculated with the M.C., 
G.C and canonical ensembles for the LDM1 and LDM2 formulae.
For details, see the text.
}
\end{figure}

Finally, we check whether our conclusions depend on the statistical ensemble employed in the calculations.
Figure \ref{fig:alphaEnsembs} shows a comparison between the values of $\alpha$ obtained with the M.C., canonical, and G.C.
ensembles, using the two LDM formulae discussed in this work.
One observes very small differences at large temperatures, {\it i.e.} for $T>4-5$~MeV, which is the temperature domain in which
these models are usually employed in the study of the nuclear multifragmentation.
The discrepancies at low temperatures between the canonical and the G.C. ensembles are due to finite size effects, since these
systems are far from the thermodynamical limit \cite{BettyPhysRep2005}.
The deviations are more important in the case of the M.C. calculations at low temperatures where the excitation
energy of the system is very small and the strict energy conservation constraint plays an important role in determining the most
important partitions.
These results suggest that our conclusions should not be affected by the use of different ensembles in the temperature domain
which is relevant to the multifragment emission.

\section{\label{sec:conclusions}Concluding remarks\protect}
We have demonstrated that the isoscaling parameter $\alpha$ is not sensitive to the Coulomb and the surface
terms of the nuclear binding energy in the temperature domain in which the multifragment emission is expected to take place, although they
are very important in the vanishing temperature limit.
The symmetry energy dictates the behavior of $\alpha$ for $T\gtrsim 5.0$~MeV, despite the fact that the simple analytical formulae that
relate $\alpha$ to $C_{\rm sym}$, Eqs.\ (\ref{eq:isoAlpha}) or (\ref{eq:alphaT}), are not valid in this temperature domain.
The weak dependence of $\alpha$ on the entropic terms of the internal excitation energy of the fragments suggests that,
at high temperatures, $\alpha$ is essentially governed by the symmetry energy and $T$, although its actual functional dependence
is rather complex in this regime.
More precisely, we found that $\alpha$ is given by an unknown function $f$ which, for $T\gtrsim 5.0$~MeV, 
is not sensitive to the Coulomb and surface terms, {\it i.e.} $\alpha=f(C_{\rm sym},(Z_1/A_1)^2-(Z_2/A_2)^2,T)$.
Furthermore, we also found that $\alpha\ne 4C_{\rm sym}[(Z_1/A_1)^2-(Z_2/A_2)^2]/T$, except for $T\rightarrow 0$.
Despite this fact, since $\alpha=f(C_{\rm sym},(Z_1/A_1)^2-(Z_2/A_2)^2,T)$ in the temperature domain at which the
multifragment emission is expected to occur, it suggests that $\alpha$ is indeed a good probe for the symmetry energy coefficient but
interpretations based on the isoscaling analysis should be taken with care.
We also found that, for $T\gtrsim 4$ or 5~MeV, the isoscaling analysis is not sensitive to the statistical ensemble employed.

\begin{acknowledgments}
We would like to acknowledge CNPq, FAPERJ, PROSUL, and the PRONEX program under contract 
No E-26/171.528/2006, for partial financial support.
This work was supported in part by the National Science Foundation under Grant
Nos.\ PHY-0606007 and INT-0228058.
AWS is supported by NSF grant 04-56903.
\end{acknowledgments}

\bibliography{isotemp}

\end{document}